\documentclass[%
 reprint,
superscriptaddress,
 amsmath,amssymb,
 aps,
 pre,
floatfix,
showpacs,
nofootinbib
]{revtex4-1}

\usepackage{graphicx}

\usepackage{graphicx}
\usepackage{dcolumn}
\usepackage{bm}
\usepackage{mathbbol}
\usepackage{multirow}
\usepackage{scalerel}
\usepackage{relsize}
\usepackage[hidelinks]{hyperref}
\usepackage{xcolor}
\hypersetup{colorlinks,linkcolor={blue},citecolor={blue},urlcolor={black}}

\usepackage{comment}

\usepackage{tikz}
\usetikzlibrary{patterns,snakes}

\usepackage{xspace}
\newcommand{\elabel}[1]{\label{eq:#1}}

\newcommand{\Eref}[1]{Eq.~(\ref{eq:#1})}

\newcommand{\flabel}[1]{\label{fig:#1}}

\newcommand{\Fref}[1]{Figure~\ref{fig:#1}}

\newcommand{\tlabel}[1]{\label{tab:#1}}
\newcommand{\Tref}[1]{Tab.~\ref{tab:#1}}

\newcommand{\latin}[1]{{\it #1}}
\newcommand{\ie}{\latin{i.e.}\@\xspace}

\newcommand{\zetaMFc}{\zeta_{c,\text{\scriptsize(MF)}}}

\begin{document}

\title{Critical Density of the Abelian Manna Model via a Multi-type Branching Process}
\author{Nanxin Wei}
\email{n.wei14@imperial.ac.uk}
\author{Gunnar Pruessner}
\email{g.pruessner@imperial.ac.uk}
\affiliation{Department of Mathematics, Imperial College London, London SW7 2AZ, United Kingdom}
\affiliation{Centre for Complexity Science, Imperial College London, SW7
2AZ London, United Kingdom}
\date{\today}

\begin{abstract}
A multi-type branching process is introduced to mimic the evolution of the avalanche activity and determine the critical 
density of the Abelian Manna model. This branching process incorporates partially the spatio-temporal correlations of the 
activity, which are essential for the dynamics, in particular in low dimensions. An analytical expression for
the critical density in arbitrary dimensions is derived, which significantly improves the results over mean-field theories, 
as confirmed by comparison to the literature on numerical estimates from simulations. 
The method can easily be extended to lattices and dynamics other than those studied in the present work.
\end{abstract}

\maketitle

\section{Introduction}
The Manna Model \cite{manna1991two} is the prototypical stochastic sandpile model proposed for self-organized criticality (SOC) 
\cite{bak1988self}. It was reformulated by Dhar to make it Abelian \cite{dhar1999abelian}.
The resulting Abelian Manna model (AMM) and its variants have been studied extensively numerically and analytically
\cite{ben1996universality,dhar1999some,dickman2002n,basu2014self,wiese2016coherent,willis2018spatio}.
Numerical simulations have established that a range of other models belong to the same universality class \cite[p. 178]{Pruessner2012self},
in particular the Oslo Model
\cite{ChristensenETAL:1996,PaczuskiBoettcher:1996} and the conserved
lattice gas \cite{RossiPastor-SatorrasVespignani:2000,Jensen:1990}. The
stationary density of the AMM has been estimated with very high precision 
on hypercubic lattices of dimensions $d=1$ to $d=5$
\cite{lubeck2000moment,dickman2001critical,HuynhPruessnerChew:2011,HuynhPruessner:2012b,willis2018spatio}.
Yet, theoretical understanding 
of the Manna Model is far from complete. In a mean-field theory which ignores all spatial-temporal correlations, 
the avalanches can be naturally perceived as a binary branching process (BP) \cite{alstrom1988mean,garcia1994dimension} 
with branching ratio $\sigma$ twice the particle density $\zeta$ as a mean field. At stationarity, the macroscopic dynamics of driving and 
dissipation of particles \emph{self-organises} the branching ratio to unity, which is the critical value, $\sigma_c=1$, \ie 
the branching ratio above which a finite fraction of realisations branches indefinitely.
The mean-field value of the critical 
density is therefore $\zetaMFc = \sigma_c/2=1/2$ regardless of the dimension of the system \cite{ZapperiLauritsenStanley:1995}. However, numerical 
findings have placed the critical value $\zeta_{c}$ of the density
clearly above $1/2$ in \emph{any} dimension studied \cite{Pruessner2012self,willis2018spatio}, 
suggesting that the spatial correlations ignored by the mean-field theory are significant. 
Here, we provide a theoretical characterisation of $\zeta_{c}$ in a general 
setting through a mapping to a multi-type branching process (MTBP), with  a
simple closed-form approximation systematically 
improving on the mean-field prediction. Our method incorporates 
only short-ranged 
correlations during the avalanche and highlights the role that particles 
conservation plays in regulating 
activity.
While still ignoring 
correlations in the initial state \cite{BasuETAL:2012,HexnerLevine:2015,willis2018spatio}, we show that 
taking into account even only some of the 
correlations arising in the activity modifies significantly the estimate of the critical density. One may 
hope that our findings can be reproduced to leading order in a suitable field theory.

In the following, we first introduce the Abelian Manna Model and its
(approximate) mapping to
the multi-type branching process mimicking (some of) its dynamics. We
then demonstrate how the critical density of the AMM can be
extracted from the branching process and conclude with a brief
discussion of the results.

\section{The Abelian Manna Model}
To facilitate the following discussion, we reproduce the definition of
the AMM:
The AMM is normally studied on a $d$-dimensional hypercubic lattice, but extensions to arbitrary graphs are straight forward.
Each site carries a non-negative number of particles, which we refer to as the \emph{occupation number}. 
A site that carries no particle is said to be \emph{empty}, otherwise it
is \emph{occupied} by at least one particle.
A site carrying less than two particles is \emph{stable}, 
otherwise it is \emph{active}.
If all sites in a lattice are stable, 
the system is said to be \emph{quiescent}, as it does not evolve by its internal dynamics.
If the lattice has $N$ sites, the number of such states is $2^N$.
Particles are added to the lattice by an \emph{external drive}. If such
an externally added particle arrives at a site that is occupied by a
particle already, an avalanche ensues as follows: Every site that
carries more than one particle \emph{topples} by moving  two of them to
randomly and independently chosen nearest neighbours, thereby
\emph{charging} them with particles.
This might trigger a toppling in turn. The totality of all topplings in response to a single particle added by the external drive is called an avalanche.

We will refer to the evolution from toppling to toppling as the
\emph{microscopic timescale}, as opposed to the macroscopic timescale
of the evolution of quiescent states.
The evolution from one quiescent state to another quiescent state by adding a particle at a site and letting an avalanche complete
is a Markov process. 
Because of the finiteness of the state space of quiescent configurations and 
assuming accessibility of all states (but see \citep{WillisPruessner:2018}), 
the probability to find the AMM in a particular quiescent state 
approaches a unique, stationary, strictly positive value. The analysis below is 
concerned solely with the the stationary state of the AMM.

For the discussion below we require the notion of occupation number 
\emph{pre-toppling} and 
occupation number \emph{post-toppling}. The former refers to the number of particles
at a site prior to its possible toppling, the latter refers to the
number of particles
at a site immediately after possibly shedding (a multiple of) two particles 
and yet prior to
it receiving particles from any other site. Committing a slight abuse of terminology,
we will refer to pre- and post-toppling occupation numbers even when the site 
is stable.

Of particular importance to the following consideration is the occupation density $\zeta$,
that 
the expected total number of particles divided by the number of sites.

\subsection{The multi-type branching process}
One paradigm of the AMM and SOC in general is the 
(binary) branching process \cite{Harris:1963}. 
The population of that process at any given time is thought to represent 
those sites that become active as a result of receiving a
particle.
As they topple, the particles arriving at nearest neighbouring stable
sites might
activate those, depending on whether they were previously occupied
by a particle or not. 
Any empty site that is charged with only one particle becomes
occupied but remains
stable.
Any stable site charged with two particles is
guaranteed to become active. 
As active sites are sparse, they are rarely charged. However, the
Abelian property of the AMM means that the arrival of one additional particle at a site leads to
a further toppling only if the parity of its occupation number is odd.
If two particles arrive at a site, its parity will not change, but the site is
bound to topple (once more).

If a neighbouring site becomes active in response to a charge,
this corresponds in the branching process to an offspring in the next
generation. If two such offspring are generated, this corresponds to a branching that increases the 
population size. If a neighbouring site turns from empty to occupied, no
offspring is produced. 


The spatio-temporal 
evolution of an avalanche may 
thus be thought of as a branching process embedded in space and
with strong correlations of branching and extinction events 
as the lattice occupation dictates whether and where these events take place. 
Ignoring 
the lattice and the history of previous and ongoing avalanches, 
one is left with a plain
binary branching processes, as it is commonly used to cast the AMM
and SOC models generally in a mean field theory \cite{ZapperiLauritsenStanley:1995,Luebeck:2004,JuanicoMonterolaSaloma:2007a,BonachelaMunoz:2009}. Field
theoretic treatments of any such processes always involve branching 
as a basic underlying process \cite{vespignani1998driving,RamascoMunozdaSilvaSantos:2004}.

In general branching processes, the branching ratio is exactly unity
at the critical point of the process, above which the probability 
of sustaining a finite population size indefinitely is strictly positive.
We therefore identify the critical 
point of the branching process with that of the lattice model.


\begin{table*}[]
\renewcommand{\arraystretch}{1.5}
\begin{tabular}{l | c | c | c | c}
\hline
dimension              & 1 & 2 & 3 & 5 \\
\hline
$\zeta_{c}$ (numerically) 	
&0.94882(1) \cite{willis2018spatio}
&0.7170(4)\cite{HuynhPruessnerChew:2011}
&0.622325(1) \cite{HuynhPruessner:2012b} 
&0.559780(5) \cite{willis2018spatio}	\\
$\zeta_{c}$ (present work)      &\hspace{8pt}0.750\hspace{8pt}  	&\hspace{8pt}0.625\hspace{8pt}   
	&\hspace{8pt}0.583\hspace{8pt}   &\hspace{8pt}0.550\hspace{8pt}   \\
\hline 
\multicolumn{5}{l}{} \vspace{-10pt}
\end{tabular} 
\caption{\tlabel{NumericsVSAnalytics}
The theoretical estimate of the critical density $\zeta_c$ 
in the $d$-dimensional AMM 
derived here
compared to the numerical values reported in the literature
for the stationary density of the Abelian Manna Model.
}
\end{table*}

\begin{figure}
\begin{tikzpicture}[scale=0.5]
\draw[ultra thick] (-1.2,0) -- (2.2,0) node[below,pos=0.5] {$\alpha_0$};
\draw[ultra thick] (-1,0) -- (-1,0.2);
\draw[ultra thick] (0,0) -- (0,0.2);
\draw[ultra thick] (1,0) -- (1,0.2);
\draw[ultra thick] (2,0) -- (2,0.2);
\draw[thick] (0,0) rectangle (1,1);
\draw[thick] (0,1) rectangle (1,2);
\draw[thick] (0,2) rectangle (1,3);
\end{tikzpicture}
$\qquad$
\begin{tikzpicture}[scale=0.5]
\draw[ultra thick] (-1.2,0) -- (2.2,0) node[below,pos=0.5] {$\alpha_1$};
\draw[ultra thick] (-1,0) -- (-1,0.2);
\draw[ultra thick] (0,0) -- (0,0.2);
\draw[ultra thick] (1,0) -- (1,0.2);
\draw[ultra thick] (2,0) -- (2,0.2);
\draw[thick] (-1,0) rectangle (0,1);
\draw[thick] (0,0) rectangle (1,1);
\draw[thick] (0,1) rectangle (1,2);
\draw[thick] (0,2) rectangle (1,3);
\end{tikzpicture}
$\qquad$
\begin{tikzpicture}[scale=0.5]
\draw[ultra thick] (-1.2,0) -- (2.2,0) node[below,pos=0.5] {$\alpha_2$};
\draw[ultra thick] (-1,0) -- (-1,0.2);
\draw[ultra thick] (0,0) -- (0,0.2);
\draw[ultra thick] (1,0) -- (1,0.2);
\draw[ultra thick] (2,0) -- (2,0.2);
\draw[thick] (-1,0) rectangle (0,1);
\draw[thick] (0,0) rectangle (1,1);
\draw[thick] (0,1) rectangle (1,2);
\draw[thick] (0,2) rectangle (1,3);
\draw[thick] (1,0) rectangle (2,1);
\end{tikzpicture}

\begin{tikzpicture}[scale=0.5]
\draw[ultra thick] (-1.2,0) -- (2.2,0) node[below,pos=0.5] {$\beta_0$};
\draw[ultra thick] (-1,0) -- (-1,0.2);
\draw[ultra thick] (0,0) -- (0,0.2);
\draw[ultra thick] (1,0) -- (1,0.2);
\draw[ultra thick] (2,0) -- (2,0.2);
\draw[thick] (0,0) rectangle (1,1);
\draw[thick] (0,1) rectangle (1,2);
\end{tikzpicture}
$\qquad$
\begin{tikzpicture}[scale=0.5]
\draw[ultra thick] (-1.2,0) -- (2.2,0) node[below,pos=0.5] {$\beta_1$};
\draw[ultra thick] (-1,0) -- (-1,0.2);
\draw[ultra thick] (0,0) -- (0,0.2);
\draw[ultra thick] (1,0) -- (1,0.2);
\draw[ultra thick] (2,0) -- (2,0.2);
\draw[thick] (-1,0) rectangle (0,1);
\draw[thick] (0,0) rectangle (1,1);
\draw[thick] (0,1) rectangle (1,2);
\end{tikzpicture}
$\qquad$
\begin{tikzpicture}[scale=0.5]
\draw[ultra thick] (-1.2,0) -- (2.2,0) node[below,pos=0.5] {$\beta_2$};
\draw[ultra thick] (-1,0) -- (-1,0.2);
\draw[ultra thick] (0,0) -- (0,0.2);
\draw[ultra thick] (1,0) -- (1,0.2);
\draw[ultra thick] (2,0) -- (2,0.2);
\draw[thick] (-1,0) rectangle (0,1);
\draw[thick] (0,0) rectangle (1,1);
\draw[thick] (0,1) rectangle (1,2);
\draw[thick] (1,0) rectangle (2,1);
\end{tikzpicture}
\caption{\flabel{active_motives}
The six different \emph{active motives} of the one-dimensional Abelian Manna Model. 
The central site 
carries either
three
($\alpha_i$ for $i=0,1,2$) or 
two
particles ($\beta_i$ for $i=0,1,2$)
and
is to topple in the next 
microscopic time step. 
The occupation number of sites neighbouring the central 
site are representative only in as far as their parity 
is concerned, with $i$ indicating the number of sites that
carry an odd number. Their occupation number is \emph{post-toppling} (after
\emph{they} may have toppled themselves, but prior to the central site toppling),
whereas the central site is shown \emph{pre-topping} (before it topples).
The configurations differ in whether neighbouring sites may or may not topple themselves due to  
being charged by the toppling site.
Although we may picture the configurations being situated 
on a one-dimensional lattice, their spatial orientation 
is irrelevant and so we do not distinguish left and right 
neighbours.
}
\end{figure}
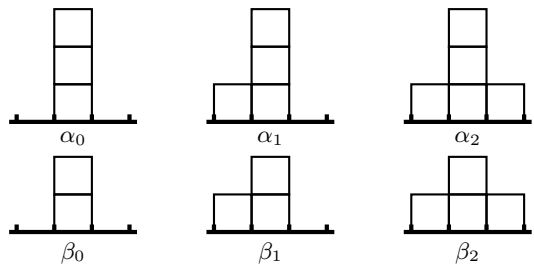

The multi-type branching process considered in the following is based on
a mapping of the types (or species) of the branching process to the active \emph{motives} of the AMM, as shown in
\Fref{active_motives} for the one-dimensional case.
These motives indicate the occupation number of the central site pre-toppling 
(\ie prior to the central site toppling)
and the occupation of the neighbouring sites post-toppling 
(\ie after they may have toppled themselves, which leaves their parity unchanged, 
but prior to the central site toppling).
Defining the motives this way, we can disregard toppling sites charging active neighbouring sites.
We effectively  
keep track
only of the change of parity of the neighbouring sites, which is due to charges they receive but does
not change when they topple themselves.
Within a time step 
in the multi-type branching process all currently active motives undergo toppling, which 
corresponds to
parallel updating on the microscopic time scale of the AMM.

The species labels 
in \Fref{active_motives}
of the form $\sigma_i$ characterise the configuration as follows: 
$\sigma=\alpha$ indicates that the active site carries three particles, $\sigma=\beta$ indicates that it carries two. The index $i$ indicates the number of neighbouring sites carrying a single particle post-toppling. 
In general, $i\in\{0,\ldots,q\}$ where $q$ is the coordination number. Henceforth, we restrict ourselves
to regular lattices with constant coordination number $q$. These may be
thought of as 
$d$-dimensional hypercubic lattices with $q=2d$.

At any point during the evolution of
an avalanche, those sites whose occupation
information is not captured by the active motives are occupied independently with
density $\zeta$. 
The active sites we consider carry only ever two or three particles, \ie 
we do not keep track of multiple topplings.
In \Fref{toppling_example} we illustrate how the toppling of motive $\alpha_1$ in one
dimension gives rise to 
the motive $\beta_2$.

In the interest of clarity, we summarise our key assumptions:
\emph{(i)} In each timestep during 
an avalanche, the substrate sites (sites whose occupation information is
not captured in the active motives) are assumed to be occupied 
independently with probability $\zeta$, which is a fixed model parameter.
This is where we ignore correlations.
\emph{(ii)} No 
occupation number post-toppling exceeds unity and no
active 
site carries more than three particles.
This is a significant restriction only in one dimension, where multiple
toppling is known to play a significant role \cite{nonotePaczuskiBassler:2000}.
\emph{(iii)} No site 
receives particles toppling from different sites simultaneously. 
\emph{(iv)} All sites are considered bulk sites, \ie there is no boundary. Each site has therefore the same number $q$ of 
neighbours.

\newcommand{\Bin}{\operatorname{Bin}}

\renewcommand{\wp}{w.p.\@\xspace}

The time in the branching process progresses by all individuals
attempting branching
in each parallel timestep, which corresponds to the microscopic time in the AMM.
The branching itself mimics the toppling dynamics:
In each toppling two particles are redistributed to the same neighbour 
with probability (\wp) $1/q$ and to different neighbours \wp $1-1/q$.
For a configuration of type $\sigma_j$, a randomly chosen neighbour 
of the active, toppling site is occupied (has odd parity)
\wp $j/q$. The active site itself
will be left occupied if $\sigma =\alpha$ and empty if $\sigma=\beta$.
The next nearest neighbours of any active site are treated as substrate sites, occupied 
\wp $\zeta$ and empty \wp $(1-\zeta)$. 
An illustrative example of 
a branching path on a one-dimensional lattice is shown in \Fref{toppling_example},
where the motive
$\alpha_{1}$ is shown to turn into $\beta_{2}$ \wp $\zeta/4$.
The probabilities
of all branching paths on regular
lattices with constant $q$ are listed in \Tref{BranchingProbabilities}. 
The multi-type branching process is initialised 
by a single node of type $\beta_{j}$, which is drawn with probability 
\begin{equation}\elabel{def_Bin}
    \Bin(j,q;\zeta) = {q \choose j}\  \zeta^j (1-\zeta)^{q-j},
\end{equation}
reflecting 
the fact that an avalanche in the AMM is initialised by a single 
site driven to active by the external drive from a quiescent configuration.
To make the expressions below well-defined, we define $\Bin(j,q;\zeta)=0$ for 
$j>q$.

\begin{figure}[h] 	
\begin{tikzpicture}[scale=0.5]
\begin{scope}[xshift=-0.5cm]
\draw[ultra thick] (-1.2,0) -- (2.2,0) node[below,pos=0.5] {$ $};
\draw [
    thick,
    decoration={
        brace,
        mirror,
        raise=0.1cm
    },
    decorate
] (-1,0) -- (2,0) 
node [pos=0.5,anchor=north,yshift=-0.55cm] {$\alpha_1$};

\draw[ultra thick] (2.2,0) -- (3.2,0);
\draw[ultra thick] (-1,0) -- (-1,0.2);
\draw[ultra thick] (0,0) -- (0,0.2);
\draw[ultra thick] (1,0) -- (1,0.2);
\draw[ultra thick] (2,0) -- (2,0.2);
\draw[ultra thick] (3,0) -- (3,0.2);
\draw[thick] (-1,0) rectangle (0,1);
\draw[thick] (0,0) rectangle (1,1);
\draw[thick] (0,1) rectangle (1,2);
\draw[thick] (0,2) rectangle (1,3);
\draw[thick,dotted] (2,0) rectangle (3,1);
\draw[] (3.85,0.75) node[] {occ.};
\draw[] (4.2,0.25) node[] {w.p. $\!\!\zeta$};

\draw[->] (1,1.5) to[out=0,in=90] (1.5,1);
\draw[] (2.5,1.6) node {w.p. $\!\!1/2$};
\draw[->] (1,2.5) to[out=0,in=90] (1.5,2);
\draw[] (2.5,2.6) node {w.p. $\!\!1/2$};

\end{scope}

\begin{scope}[xshift=0.3cm]
\draw[ultra thick,->] (4,1.7) -- (8,1.7) node[above,pos=0.5]{w.p. $\zeta/4$};  
\end{scope}

\begin{scope}[xshift=10.5cm]
\draw[ultra thick] (-1.2,0) -- (3.2,0) node[below,pos=0.5] {$ $};
\draw [
    thick,
    decoration={
        brace,
        mirror,
        raise=0.1cm
    },
    decorate
] (0,0) -- (3,0) 
node [pos=0.5,anchor=north,yshift=-0.55cm] {$\beta_2$}; 

\draw[ultra thick] (2.2,0) -- (3.2,0);
\draw[ultra thick] (-1,0) -- (-1,0.2);
\draw[ultra thick] (0,0) -- (0,0.2);
\draw[ultra thick] (1,0) -- (1,0.2);
\draw[ultra thick] (2,0) -- (2,0.2);
\draw[ultra thick] (3,0) -- (3,0.2);
\draw[thick] (-1,0) rectangle (0,1);
\draw[thick] (0,0) rectangle (1,1);
\draw[thick] (1,1) rectangle (2,2);
\draw[thick] (1,0) rectangle (2,1);
\draw[thick] (2,0) rectangle (3,1);

\end{scope}

\end{tikzpicture}
   \caption{Example of a toppling on a one-dimensional lattice. 
   The initial state 
   $\alpha_{1}$ goes over into state $\beta_{2}$ 
   with probability (\wp) $\zeta/4$, which is the joint probability of three 
   independent events: The independent toppling of two particles to one
   side (\wp $1/4$) and the occupation of a next-nearest neighbouring 
   site (\wp $\zeta$). The latter assumption ignores spatial correlations.}
\flabel{toppling_example} 
\end{figure}
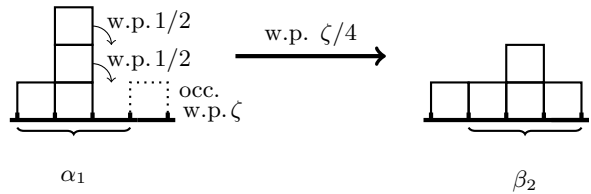

\begin{table}[]
\renewcommand{\arraystretch}{1.5}
\begin{tabular}{|rr|r|}
\hline
parent&\multicolumn{1}{|c|}{branching probability}&offspring\\
\hline
\multirow{4}{*}{$\alpha_{j}\left\{\begin{array}{l}
\\
\\
\\
\!
\end{array}\right.\hspace*{-3.5mm}
$}
& $\frac{j}{q^2}           \ \Bin(k,q-1;\zeta)$ & $\alpha_{k+1}$\\
& $\frac{(2j+1)(q-j)}{q^2}     \ \Bin(k,q-1;\zeta)$ & $\beta_{k+1}$\\
& $\frac{j(j-1)}{q^2}      \ \Bin(k,q-1;\zeta)$ & $2\beta_{k+1}$\\
& $\frac{(q-j)(q-j-1)}{q^2}\ \Bin(k,q-1;\zeta)$ & $\varnothing$\\
\hline
\multirow{4}{*}{$\beta_{j}\left\{\begin{array}{l}
\\
\\
\\
\!
\end{array}\right.\hspace*{-3.5mm}
$}
& $\frac{j}{q^2}           \ \Bin(k,q-1;\zeta)$ & $\alpha_{k}$\\
& $\frac{(2j+1)(q-j)}{q^2}     \ \Bin(k,q-1;\zeta)$ & $\beta_{k}$\\
& $\frac{j(j-1)}{q^2}      \ \Bin(k,q-1;\zeta)$ & $2 \beta_{k}$\\
& $\frac{(q-j)(q-j-1)}{q^2}\ \Bin(k,q-1;\zeta)$ & $\varnothing$\\
\hline
\end{tabular} 
\caption{
Branching probabilities of 
active motives on
a regular lattice with constant coordination number $q$.
The only form of branching occurs
is when 
two independent copies of $\beta_{k}$ are generated,
indicated by $2\beta_{k}$. The symbol $\varnothing$ indicates that no
offspring is produced. 
\tlabel{BranchingProbabilities}
}\end{table}

\section{Critical density}
The multi-type branching process defined above approximates the
population dynamics of the activity in an avalanche of the AMM on an infinite lattice. 
Activity performs a (branching) 
random walk on the lattice, as active sites 
topple and produce active offspring sites \cite{Pruessner_aves:2013}.
For avalanches on finite lattices with open boundaries, when the density is subcritical, the activity is expected to 
extinguish before particles reach any of the boundaries, 
and as a result, the occupation density $\zeta$ 
increases under the external drive.
When supercritical, the activity with large probability persists until incurring dissipation at the boundaries which decreases the 
density accordingly \cite{nonotePaczuskiBassler:2000}. On the other hand, large $\zeta$ will generally lead to larger avalanches, and small $\zeta$ to small avalanches.
Nevertheless, under this apparent
self-organisation \cite{DickmanVespignaniZapperi:1998,PetersPruessner:2006}, the fluctuations of $\zeta$ decrease with 
system size and in the thermodynamic limit, the stationary
density approaches
a particular value generally referred to as the critical density
(even when there may be more than one, \cite{FeyLevineWilson:2010b}).
We identify the critical density as the 
smallest density $\zeta_c$
at which 
the multi-type branching process has a finite probability to evolve 
forever, \ie its critical point, when the branching ratio is unity.

\newcommand{\Mmatrix}{\mathbf{M}}

To find the critical
point of the multi-type branching process, it suffices 
to determine the density $\zeta_c$ when 
the largest eigenvalue $\lambda_1$ of the mean offspring matrix 
$\Mmatrix$ introduced below is unity 
\citep[][Theorem 2, V.3]{athreya2012branching}.

The mean offspring matrix is the matrix $\Mmatrix=\{m_{\sigma,\tau}\}$,
with $m_{\sigma,\tau}$ the mean number of offspring of type $\tau$ produced
as an individual of type $\sigma$ undergoes multi-type branching, \ie
an update. The types $\sigma$ and $\tau$ are any of the states $\alpha_j$ and 
$\beta_j$ with $j\in\{0,1,\ldots,q\}$,
as exemplified in
\Fref{active_motives}.
The individual elements $m_{\sigma,\tau}$ 
of the matrix are easily determined 
from \Tref{BranchingProbabilities}, 
\begin{subequations}
\begin{align}
    m_{\alpha_j,\alpha_0} &= 0\\
    m_{\alpha_j,\alpha_k} &= \frac{j}{q^2} \Bin(k-1,q-1;\zeta) \text{ for } k>0\\
    m_{\alpha_j,\beta_0} &= 0\\
    m_{\alpha_j,\beta_k}  &= \frac{(2j+1)q-3j}{q^2} \Bin(k-1,q-1;\zeta) \text{ for } k>0\\
    m_{\beta_j,\alpha_k} &= \frac{j}{q^2} \Bin(k,q-1;\zeta)\\
    m_{\beta_j,\beta_k}  &= \frac{(2j+1)q-3j}{q^2} \Bin(k,q-1;\zeta).
\end{align}
\end{subequations}
Because 
$m_{\alpha_j,\alpha_{k+1}}=m_{\beta_j,\alpha_k}$ 
and
$m_{\alpha_j,\beta_{k+1}}=m_{\beta_j,\beta_k}$, 
the matrix $\Mmatrix$ has some very convenient symmetries, which,
after ordering states according to $(\alpha_0,\ldots,\alpha_q,\beta_0,\ldots,\beta_q)$
may be written as
\begin{equation}
\Mmatrix=\left(
\begin{array}{c|c}
\underline{0}_{q+1},\underline{a}\bigotimes\underline{B}\ & \underline{0}_{q+1},\underline{b}\bigotimes\underline{B} \\ \hline
\underline{a}\bigotimes\underline{B},\underline{0}_{n+1} & \underline{b}\bigotimes\underline{B},\underline{0}_{q+1}
\end{array}\right)
\end{equation}
with vectors
\begin{align}
\underline{a}=(a_{j})_{0\leqslant j\leqslant q} &\quad \text{with }\quad a_{j}=j/q^{2} \\
\underline{b}=(b_{j})_{0\leqslant j\leqslant q} &\quad \text{with }\quad b_{j}=(2j+1)/q-3j/q^{2}\\
\underline{B}=(B_{k})_{0\leqslant k\leqslant q-1} &\quad \text{with }\quad B_{k}=\Bin(k,q-1;\zeta) 
\end{align}
and $\underline{0}_{q+1}$ is a column of $q+1$ zeroes. For example, the
matrix for the one-dimensional AMM
is 
\[
\Mmatrix=
\left(
\begin{array}{rrrrrr}

    0 & 0 & 0 &
    0 & \frac{1}{2} B_{0} &  \frac{1}{2} B_{1} \\[6pt]
    0 & \frac{1}{4} B_0 & \frac{1}{4} B_{1} &
    0 & \frac{3}{4} B_{0} & \frac{3}{4} B_{1} \\[6pt]
    0 & \frac{1}{2} B_0 & \frac{1}{2} B_{1} &
    0 & B_0 & B_1 \\[6pt]
    0 & 0 & 0 &
    \frac{1}{2} B_0 & \frac{1}{2} B_1 & 0 \\[6pt]
    \frac{1}{4} B_0 & \frac{1}{4} B_1 & 0 &
    \frac{3}{4} B_{0} & \frac{3}{4} B_{1} & 0 \\[6pt]
    \frac{1}{2} B_0 & \frac{1}{2} B_1 & 0 &
    B_0 & B_1 & 0
\end{array}
\right) \ ,
\]
using $B_0=\Bin(0,1;\zeta)=1-\zeta$, $B_1=\Bin(1,1;\zeta)=\zeta$ and $B_2=\Bin(2,1;\zeta)=0$.

Upon factorisation, the 
characteristic polynomial of $\mathbf{M}$ obtains a surprisingly simple form:
\begin{equation}\label{eq2}
\begin{split}
&\operatorname{det}\big(\lambda\mathbf{I} - \mathbf{M}\big) = \\
&\lambda^{2q-2}\Big(\lambda^{2}-\frac{1}{q^{3}}\Big)\Big[\lambda^{2}-\frac{2(q-1)^{2}\zeta +q+1}{q^{2}}+\frac{1}{q^{3}}\Big]
.
\end{split}
\end{equation}
Since $q=2d \geqslant 2$, the largest root $\lambda_{1}$ equals unity 
if and only if $\zeta=\frac{q+1}{2q}$. It exceeds unity if and only if $\zeta$ exceeds
$\frac{q+1}{2q}$. Our estimate of the critical density 
is thus 
\begin{equation}\elabel{central_result}
\zeta_{c}=\frac{q+1}{2q}.
\end{equation}
This is the central result of the present work. 
Writing this result in a more suggestive form, with $q=2d$ for
hypercubic lattices
we obtain 
$\zeta_{c}=1/2+ 1/(4d)$, \ie the correction to the mean-field result
$\zetaMFc=1/2$ is $1/(4d)$.
\Tref{NumericsVSAnalytics} shows a 
comparison between this result 
and the numerical values found by simulations
\cite{willis2018spatio,HuynhPruessnerChew:2011,HuynhPruessner:2012b}
on lattices in dimensions $d\in\{1,2,3,5\}$. While our estimate
\Eref{central_result} underestimates $\zeta_c$ as found numerically by
about $21\%$ in one dimension, 
this deviation drops to about $2\%$ in five dimensions. We would expect that
incorporating a larger number of nearest neighbours would improve the
estimate further \cite{dickman2002n}.

\subsection{AMM on random regular graphs}
In the derivation above
the dimension $d$ of the hypercubic lattices considered enters only in as far as the
coordination number $q=2d$ is concerned. Hence the results equally apply 
to the AMM on any graph 
with fixed coordination number. 
To demonstrate this, 
we compare numerical estimates of
the critical density in random $5$-regular graphs 
\cite{bollobas2001random} ($q=5$) to our theoretical approximation. 
To avoid the complication 
of choosing sinks or dissipation sites on graphs, we adopt the fixed-energy version of the AMM \cite{VespignaniETAL:2000} in the 
simulations. For a given density, particles are initially uniformly randomly distributed on the sites 
of the graph, and a random occupied site is driven to start the avalanche. 
To determine the critical density, we estimate the
survival probability 
of the activity after many microscopic timesteps (approximately ten times the size of the graph)
and plot it against the particle density for 
different graph sizes $N$, \Fref{randomGraphNumerics}.
The numerical estimate of $\zeta_{c} \approx 0.62$ 
as the apparent onset of a finite probability of indefinite 
survival
is consistent with 
our theoretical approximate $\zeta_{c} = 1/2 + 1/q = 0.6$. 

\begin{figure}[h]
\centering
\includegraphics[width=0.4\textwidth]{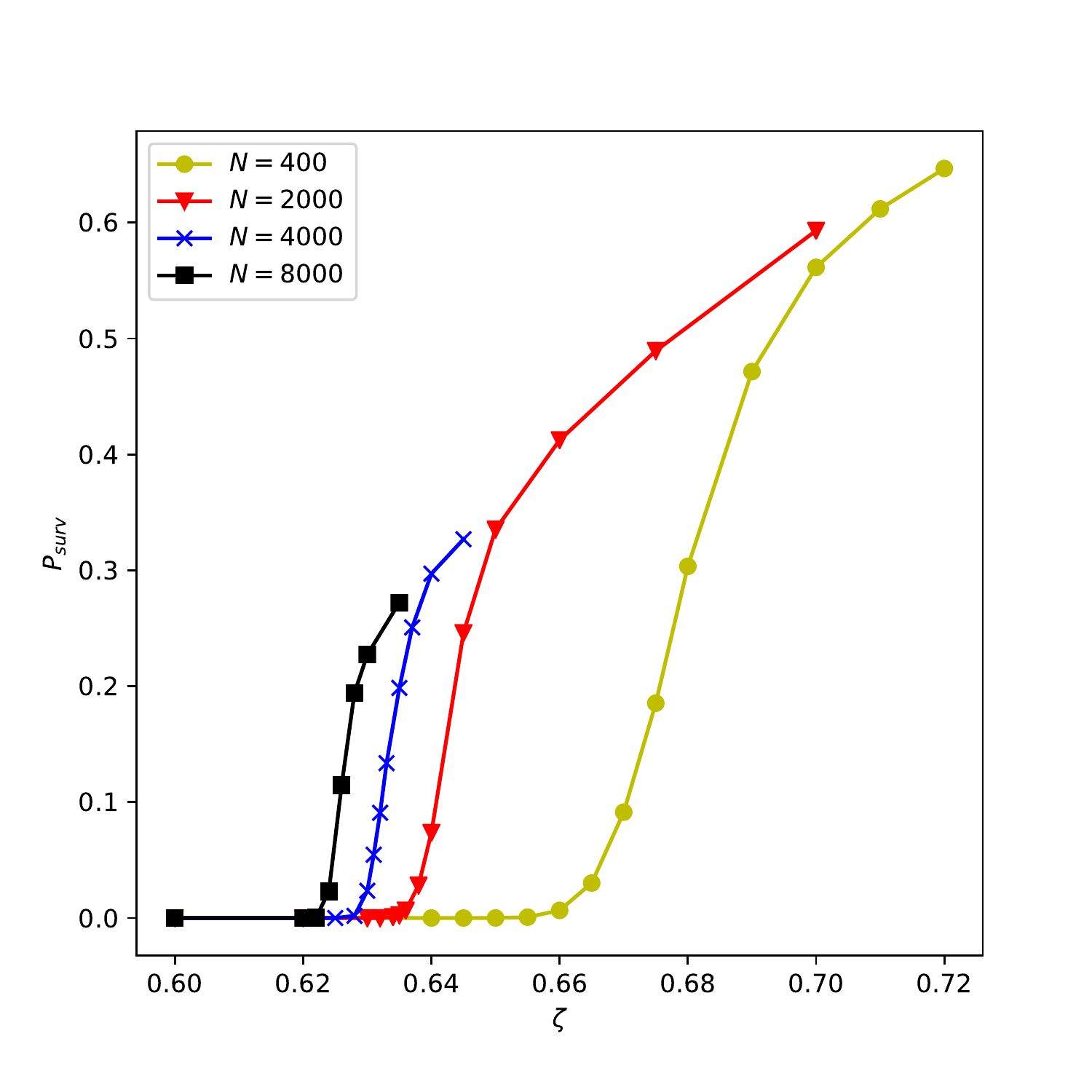}
\caption{
\flabel{randomGraphNumerics} 
Estimate of the asymptotic density in the fixed-energy 
version of the AMM on random $5$-regular graphs of different sizes.
The approximate critical point, identified as the onset of
asymptotic survival, at around $\zeta_c=0.62$ is very close
to the theoretical prediction $\zeta_{c} = 1/2 + 1/q = 0.6$.
}
\end{figure}

\section{Discussion}
In the procedure outlined 
above, we have cast the dynamics of the Abelian Manna Model in a multi-type
branching process, whose species consist 
of multiple-site 
motives of
active sites. Upon charging a singly occupied site a branching
process ensues and evolves by producing offspring according to the
density of occupied sites $\zeta$. The critical density $\zeta_c$ of the
AMM is
identified with the value of $\zeta$ when the branching process is
critical. Our main results \Eref{central_result} is in line with numerical findings in
the literature, \Tref{NumericsVSAnalytics}. The most significant
corrections are found in
low dimensions and almost perfect agreement in dimension $d=5$.

The multi-type branching process mapping we introduce keeps track only
of the parity of active sites  
and the total number of particles post-toppling 
at their neighbours. 
These few 
motives allow us to find a closed form estimate of the critical 
density on regular lattices and 
more complicated graphs. 
Compared with other theoretical methods which 
characterise the critical density 
of the AMM 
such as real-space renormalisation group 
\cite{PietroneroVespignaniZapperi:1994,lin2010renormalization} and n-site 
approximations \cite{dickman2002sandpiles,dickman2002n}, our approach 
utilises only the local topology of the underlying graph, rendering it more flexible 
and easier to generalise. 
In spirit, these active motives are 
closely related to those in the
Approximate Master Equation method \cite{gleeson2011high,gleeson2013binary}
which improves significantly on a pair approximation.
The treatment of the AMM activity here differs from that method by capturing mobile branching 
motives. To our knowledge, this has not been considered 
in the literature before. 

The main focus of the 
present work is not to improve the 
estimates of the critical density in one dimension, 
which display the most significant 
deviation from the mean-field value. 
Rather, we identify the key ingredients that 
contribute to the deviation of the critical density from the mean-field value and 
characterise the deviation analytically.
The AMM is believed to belong to the conserved 
directed percolation (CDP) universality class \cite{VespignaniETAL:2000, Rossi:2000}, 
which is different from the more general directed percolation (DP) universality class 
due to the conservation of particles in the dynamics. Through the mapping of AMM avalanches 
to a multi-type branching process, we explicitly show that the nearest-neighbour dynamical 
correlations and conservation of particles during avalanching largely capture the shift of 
the critical density away from its mean-field value, as prescribed by a simple binary 
branching process. 

Our mimicking process provides insight into the evolution of activity
in AMM avalanches. During an avalanche, activity grows (active motives branch) at the cost 
of singly-occupied sites, so that the sites receiving toppling particles are occupied with 
a probability less than their mean density. The conservation of particles and their spatial 
correlations thus lead to local suppression of branching. 
Two phenomena are ignored in our mapping of the AMM. Firstly,
the total number of particles in the 
system during avalanching may be reduced due to dissipation at open boundaries, and the 
number of singly occupied sites may decrease because of this as well as because of growing 
activity. As a result, branching is suppressed
globally, yet this effect is weak, as only a few sites
are affected \cite{WillisPruessner:2018}. 
Secondly, 
We ignore 
long-ranged anti-correlations \cite{BasuETAL:2012,HexnerLevine:2015} in the quiescent 
state of the AMM, which, however, appear to be 
rather weak albeit algebraic \cite{WillisPruessner:2018}. 
Building on the mapping we construct here, it would be interesting for
future work to establish a rigorous lower bound of the critical density in the AMM by 
associating the activity with some critical population dynamics, for example via the 
coupling method \cite{lindvall2002lectures,levin2017markov}.

\begin{acknowledgments}
The authors would like to thank Henrik Jensen and Nicolas Moloney for 
useful discussions.
\end{acknowledgments}

\bibliography{MultiTypeBranchingProcessMimickingManna}

\begin{thebibliography}{46}
\expandafter\ifx\csname natexlab\endcsname\relax\def\natexlab#1{#1}\fi
\expandafter\ifx\csname bibnamefont\endcsname\relax
  \def\bibnamefont#1{#1}\fi
\expandafter\ifx\csname bibfnamefont\endcsname\relax
  \def\bibfnamefont#1{#1}\fi
\expandafter\ifx\csname citenamefont\endcsname\relax
  \def\citenamefont#1{#1}\fi
\expandafter\ifx\csname url\endcsname\relax
  \def\url#1{\texttt{#1}}\fi
\expandafter\ifx\csname urlprefix\endcsname\relax\def\urlprefix{URL }\fi
\providecommand{\bibinfo}[2]{#2}
\providecommand{\eprint}[2][]{\url{#2}}

\bibitem[{\citenamefont{Manna}(1991)}]{manna1991two}
\bibinfo{author}{\bibfnamefont{S.}~\bibnamefont{Manna}}, \bibinfo{journal}{J.
  Phys. A} \textbf{\bibinfo{volume}{24}}, \bibinfo{pages}{L363}
  (\bibinfo{year}{1991}).

\bibitem[{\citenamefont{Bak et~al.}(1988)\citenamefont{Bak, Tang, and
  Wiesenfeld}}]{bak1988self}
\bibinfo{author}{\bibfnamefont{P.}~\bibnamefont{Bak}},
  \bibinfo{author}{\bibfnamefont{C.}~\bibnamefont{Tang}}, \bibnamefont{and}
  \bibinfo{author}{\bibfnamefont{K.}~\bibnamefont{Wiesenfeld}},
  \bibinfo{journal}{Phys. Rev. A} \textbf{\bibinfo{volume}{38}},
  \bibinfo{pages}{364} (\bibinfo{year}{1988}).

\bibitem[{\citenamefont{Dhar}(1999{\natexlab{a}})}]{dhar1999abelian}
\bibinfo{author}{\bibfnamefont{D.}~\bibnamefont{Dhar}},
  \bibinfo{journal}{Physica A} \textbf{\bibinfo{volume}{263}},
  \bibinfo{pages}{4} (\bibinfo{year}{1999}{\natexlab{a}}).

\bibitem[{\citenamefont{Ben-Hur and Biham}(1996)}]{ben1996universality}
\bibinfo{author}{\bibfnamefont{A.}~\bibnamefont{Ben-Hur}} \bibnamefont{and}
  \bibinfo{author}{\bibfnamefont{O.}~\bibnamefont{Biham}},
  \bibinfo{journal}{Phys. Rev. E} \textbf{\bibinfo{volume}{53}},
  \bibinfo{pages}{R1317} (\bibinfo{year}{1996}).

\bibitem[{\citenamefont{Dhar}(1999{\natexlab{b}})}]{dhar1999some}
\bibinfo{author}{\bibfnamefont{D.}~\bibnamefont{Dhar}},
  \bibinfo{journal}{Physica A} \textbf{\bibinfo{volume}{270}},
  \bibinfo{pages}{69} (\bibinfo{year}{1999}{\natexlab{b}}).

\bibitem[{\citenamefont{Dickman}(2002)}]{dickman2002n}
\bibinfo{author}{\bibfnamefont{R.}~\bibnamefont{Dickman}},
  \bibinfo{journal}{Phys. Rev. E} \textbf{\bibinfo{volume}{66}},
  \bibinfo{pages}{036122} (\bibinfo{year}{2002}).

\bibitem[{\citenamefont{Basu and Mohanty}(2014)}]{basu2014self}
\bibinfo{author}{\bibfnamefont{U.}~\bibnamefont{Basu}} \bibnamefont{and}
  \bibinfo{author}{\bibfnamefont{P.}~\bibnamefont{Mohanty}},
  \bibinfo{journal}{Europhys. Lett.} \textbf{\bibinfo{volume}{108}},
  \bibinfo{pages}{60002} (\bibinfo{year}{2014}).

\bibitem[{\citenamefont{Wiese}(2016)}]{wiese2016coherent}
\bibinfo{author}{\bibfnamefont{K.~J.} \bibnamefont{Wiese}},
  \bibinfo{journal}{Phys. Rev. E} \textbf{\bibinfo{volume}{93}},
  \bibinfo{pages}{042117} (\bibinfo{year}{2016}).

\bibitem[{\citenamefont{Willis and
  Pruessner}(2018{\natexlab{a}})}]{willis2018spatio}
\bibinfo{author}{\bibfnamefont{G.}~\bibnamefont{Willis}} \bibnamefont{and}
  \bibinfo{author}{\bibfnamefont{G.}~\bibnamefont{Pruessner}},
  \bibinfo{journal}{Int. J. Mod. Phys. B} \textbf{\bibinfo{volume}{32}},
  \bibinfo{pages}{1830002} (\bibinfo{year}{2018}{\natexlab{a}}).

\bibitem[{\citenamefont{Pruessner}(2012)}]{Pruessner2012self}
\bibinfo{author}{\bibfnamefont{G.}~\bibnamefont{Pruessner}},
  \emph{\bibinfo{title}{Self-organised criticality: theory, models and
  characterisation}} (\bibinfo{publisher}{Cambridge University Press},
  \bibinfo{year}{2012}).

\bibitem[{\citenamefont{Christensen et~al.}(1996)\citenamefont{Christensen,
  Corral, Frette, Feder, and J{\o}ssang}}]{ChristensenETAL:1996}
\bibinfo{author}{\bibfnamefont{K.}~\bibnamefont{Christensen}},
  \bibinfo{author}{\bibfnamefont{{\'A}.}~\bibnamefont{Corral}},
  \bibinfo{author}{\bibfnamefont{V.}~\bibnamefont{Frette}},
  \bibinfo{author}{\bibfnamefont{J.}~\bibnamefont{Feder}}, \bibnamefont{and}
  \bibinfo{author}{\bibfnamefont{T.}~\bibnamefont{J{\o}ssang}},
  \bibinfo{journal}{Phys. Rev. Lett.} \textbf{\bibinfo{volume}{77}},
  \bibinfo{pages}{107} (\bibinfo{year}{1996}).

\bibitem[{\citenamefont{Paczuski and Boettcher}(1996)}]{PaczuskiBoettcher:1996}
\bibinfo{author}{\bibfnamefont{M.}~\bibnamefont{Paczuski}} \bibnamefont{and}
  \bibinfo{author}{\bibfnamefont{S.}~\bibnamefont{Boettcher}},
  \bibinfo{journal}{Phys. Rev. Lett.} \textbf{\bibinfo{volume}{77}},
  \bibinfo{pages}{111} (\bibinfo{year}{1996}).

\bibitem[{\citenamefont{Rossi et~al.}(2000{\natexlab{a}})\citenamefont{Rossi,
  Pastor-Satorras, and Vespignani}}]{RossiPastor-SatorrasVespignani:2000}
\bibinfo{author}{\bibfnamefont{M.}~\bibnamefont{Rossi}},
  \bibinfo{author}{\bibfnamefont{R.}~\bibnamefont{Pastor-Satorras}},
  \bibnamefont{and}
  \bibinfo{author}{\bibfnamefont{A.}~\bibnamefont{Vespignani}},
  \bibinfo{journal}{Phys. Rev. Lett.} \textbf{\bibinfo{volume}{85}},
  \bibinfo{pages}{1803} (\bibinfo{year}{2000}{\natexlab{a}}).

\bibitem[{\citenamefont{Jensen}(1990)}]{Jensen:1990}
\bibinfo{author}{\bibfnamefont{H.~J.} \bibnamefont{Jensen}},
  \bibinfo{journal}{Phys. Rev. Lett.} \textbf{\bibinfo{volume}{64}},
  \bibinfo{pages}{3103} (\bibinfo{year}{1990}).

\bibitem[{\citenamefont{L{\"u}beck}(2000)}]{lubeck2000moment}
\bibinfo{author}{\bibfnamefont{S.}~\bibnamefont{L{\"u}beck}},
  \bibinfo{journal}{Phys. Rev. E} \textbf{\bibinfo{volume}{61}},
  \bibinfo{pages}{204} (\bibinfo{year}{2000}).

\bibitem[{\citenamefont{Dickman et~al.}(2001)\citenamefont{Dickman, Alava,
  Munoz, Peltola, Vespignani, and Zapperi}}]{dickman2001critical}
\bibinfo{author}{\bibfnamefont{R.}~\bibnamefont{Dickman}},
  \bibinfo{author}{\bibfnamefont{M.}~\bibnamefont{Alava}},
  \bibinfo{author}{\bibfnamefont{M.~A.} \bibnamefont{Munoz}},
  \bibinfo{author}{\bibfnamefont{J.}~\bibnamefont{Peltola}},
  \bibinfo{author}{\bibfnamefont{A.}~\bibnamefont{Vespignani}},
  \bibnamefont{and} \bibinfo{author}{\bibfnamefont{S.}~\bibnamefont{Zapperi}},
  \bibinfo{journal}{Phys. Rev. E} \textbf{\bibinfo{volume}{64}},
  \bibinfo{pages}{056104} (\bibinfo{year}{2001}).

\bibitem[{\citenamefont{Huynh et~al.}(2011)\citenamefont{Huynh, Pruessner, and
  Chew}}]{HuynhPruessnerChew:2011}
\bibinfo{author}{\bibfnamefont{H.~N.} \bibnamefont{Huynh}},
  \bibinfo{author}{\bibfnamefont{G.}~\bibnamefont{Pruessner}},
  \bibnamefont{and} \bibinfo{author}{\bibfnamefont{L.~Y.} \bibnamefont{Chew}},
  \bibinfo{journal}{J. Stat. Mech.} \textbf{\bibinfo{volume}{2011}},
  \bibinfo{pages}{P09024} (\bibinfo{year}{2011}), \eprint{arXiv:1106.0406}.

\bibitem[{\citenamefont{Huynh and Pruessner}(2012)}]{HuynhPruessner:2012b}
\bibinfo{author}{\bibfnamefont{H.~N.} \bibnamefont{Huynh}} \bibnamefont{and}
  \bibinfo{author}{\bibfnamefont{G.}~\bibnamefont{Pruessner}},
  \bibinfo{journal}{Phys. Rev. E} \textbf{\bibinfo{volume}{85}},
  \bibinfo{pages}{061133} (\bibinfo{year}{2012}), \eprint{arXiv:1201.3234}.

\bibitem[{\citenamefont{Alstr{\o}m}(1988)}]{alstrom1988mean}
\bibinfo{author}{\bibfnamefont{P.}~\bibnamefont{Alstr{\o}m}},
  \bibinfo{journal}{Phys. Rev. A} \textbf{\bibinfo{volume}{38}},
  \bibinfo{pages}{4905} (\bibinfo{year}{1988}).

\bibitem[{\citenamefont{Garc{\'\i}a-Pelayo}(1994)}]{garcia1994dimension}
\bibinfo{author}{\bibfnamefont{R.}~\bibnamefont{Garc{\'\i}a-Pelayo}},
  \bibinfo{journal}{Phys. Rev. E} \textbf{\bibinfo{volume}{49}},
  \bibinfo{pages}{4903} (\bibinfo{year}{1994}).

\bibitem[{\citenamefont{Zapperi et~al.}(1995)\citenamefont{Zapperi, Lauritsen,
  and Stanley}}]{ZapperiLauritsenStanley:1995}
\bibinfo{author}{\bibfnamefont{S.}~\bibnamefont{Zapperi}},
  \bibinfo{author}{\bibfnamefont{K.~B.} \bibnamefont{Lauritsen}},
  \bibnamefont{and} \bibinfo{author}{\bibfnamefont{H.~E.}
  \bibnamefont{Stanley}}, \bibinfo{journal}{Phys. Rev. Lett.}
  \textbf{\bibinfo{volume}{75}}, \bibinfo{pages}{4071} (\bibinfo{year}{1995}).

\bibitem[{\citenamefont{Basu et~al.}(2012)\citenamefont{Basu, Basu,
  Bondyopadhyay, Mohanty, and Hinrichsen}}]{BasuETAL:2012}
\bibinfo{author}{\bibfnamefont{M.}~\bibnamefont{Basu}},
  \bibinfo{author}{\bibfnamefont{U.}~\bibnamefont{Basu}},
  \bibinfo{author}{\bibfnamefont{S.}~\bibnamefont{Bondyopadhyay}},
  \bibinfo{author}{\bibfnamefont{P.~K.} \bibnamefont{Mohanty}},
  \bibnamefont{and}
  \bibinfo{author}{\bibfnamefont{H.}~\bibnamefont{Hinrichsen}},
  \bibinfo{journal}{Phys. Rev. Lett.} \textbf{\bibinfo{volume}{109}},
  \bibinfo{pages}{015702} (\bibinfo{year}{2012}).

\bibitem[{\citenamefont{Hexner and Levine}(2015)}]{HexnerLevine:2015}
\bibinfo{author}{\bibfnamefont{D.}~\bibnamefont{Hexner}} \bibnamefont{and}
  \bibinfo{author}{\bibfnamefont{D.}~\bibnamefont{Levine}},
  \bibinfo{journal}{Phys. Rev. Lett.} \textbf{\bibinfo{volume}{114}},
  \bibinfo{pages}{110602} (\bibinfo{year}{2015}).

\bibitem[{\citenamefont{Willis and
  Pruessner}(2018{\natexlab{b}})}]{WillisPruessner:2018}
\bibinfo{author}{\bibfnamefont{G.}~\bibnamefont{Willis}} \bibnamefont{and}
  \bibinfo{author}{\bibfnamefont{G.}~\bibnamefont{Pruessner}},
  \bibinfo{journal}{Int. J. Mod. Phys. B} \textbf{\bibinfo{volume}{32}},
  \bibinfo{pages}{1830002} (\bibinfo{year}{2018}{\natexlab{b}}),
  \eprint{arXiv:1608.00964}.

\bibitem[{\citenamefont{Harris}(1963)}]{Harris:1963}
\bibinfo{author}{\bibfnamefont{T.~E.} \bibnamefont{Harris}},
  \emph{\bibinfo{title}{The Theory of Branching Processes}}
  (\bibinfo{publisher}{Springer-Verlag}, \bibinfo{address}{Berlin, Germany},
  \bibinfo{year}{1963}).

\bibitem[{\citenamefont{L{\"u}beck}(2004)}]{Luebeck:2004}
\bibinfo{author}{\bibfnamefont{S.}~\bibnamefont{L{\"u}beck}},
  \bibinfo{journal}{Int. J. Mod. Phys. B} \textbf{\bibinfo{volume}{18}},
  \bibinfo{pages}{3977} (\bibinfo{year}{2004}).

\bibitem[{\citenamefont{Juanico et~al.}(2007)\citenamefont{Juanico, Monterola,
  and Saloma}}]{JuanicoMonterolaSaloma:2007a}
\bibinfo{author}{\bibfnamefont{D.~E.} \bibnamefont{Juanico}},
  \bibinfo{author}{\bibfnamefont{C.}~\bibnamefont{Monterola}},
  \bibnamefont{and} \bibinfo{author}{\bibfnamefont{C.}~\bibnamefont{Saloma}},
  \bibinfo{journal}{New J. Phys.} \textbf{\bibinfo{volume}{9}},
  \bibinfo{eid}{92} (pages~\bibinfo{numpages}{18}) (\bibinfo{year}{2007}).

\bibitem[{\citenamefont{Bonachela and Mu{\~n}oz}(2009)}]{BonachelaMunoz:2009}
\bibinfo{author}{\bibfnamefont{J.~A.} \bibnamefont{Bonachela}}
  \bibnamefont{and} \bibinfo{author}{\bibfnamefont{M.~A.}
  \bibnamefont{Mu{\~n}oz}}, \bibinfo{journal}{J. Stat. Mech.}
  \textbf{\bibinfo{volume}{2009}}, \bibinfo{eid}{P09009}
  (pages~\bibinfo{numpages}{37}) (\bibinfo{year}{2009}).

\bibitem[{\citenamefont{Vespignani et~al.}(1998)\citenamefont{Vespignani,
  Dickman, Mu{\~n}oz, and Zapperi}}]{vespignani1998driving}
\bibinfo{author}{\bibfnamefont{A.}~\bibnamefont{Vespignani}},
  \bibinfo{author}{\bibfnamefont{R.}~\bibnamefont{Dickman}},
  \bibinfo{author}{\bibfnamefont{M.~A.} \bibnamefont{Mu{\~n}oz}},
  \bibnamefont{and} \bibinfo{author}{\bibfnamefont{S.}~\bibnamefont{Zapperi}},
  \bibinfo{journal}{Phys. Rev. Lett.} \textbf{\bibinfo{volume}{81}},
  \bibinfo{pages}{5676} (\bibinfo{year}{1998}).

\bibitem[{\citenamefont{Ramasco et~al.}(2004)\citenamefont{Ramasco, Mu{\~n}oz,
  and da~Silva~Santos}}]{RamascoMunozdaSilvaSantos:2004}
\bibinfo{author}{\bibfnamefont{J.~J.} \bibnamefont{Ramasco}},
  \bibinfo{author}{\bibfnamefont{M.~A.} \bibnamefont{Mu{\~n}oz}},
  \bibnamefont{and} \bibinfo{author}{\bibfnamefont{C.~A.}
  \bibnamefont{da~Silva~Santos}}, \bibinfo{journal}{Phys. Rev. E}
  \textbf{\bibinfo{volume}{69}}, \bibinfo{eid}{045105(R)}
  (pages~\bibinfo{numpages}{4}) (\bibinfo{year}{2004}).

\bibitem[{\citenamefont{Paczuski and
  Bassler}(2000)}]{nonotePaczuskiBassler:2000}
\bibinfo{author}{\bibfnamefont{M.}~\bibnamefont{Paczuski}} \bibnamefont{and}
  \bibinfo{author}{\bibfnamefont{K.~E.} \bibnamefont{Bassler}}
  (\bibinfo{year}{2000}), \eprint{arXiv:cond-mat/0005340v2}.

\bibitem[{\citenamefont{Pruessner}(2013)}]{Pruessner_aves:2013}
\bibinfo{author}{\bibfnamefont{G.}~\bibnamefont{Pruessner}},
  \bibinfo{journal}{Int. J. Mod. Phys. B} \textbf{\bibinfo{volume}{27}},
  \bibinfo{pages}{1350009} (\bibinfo{year}{2013}), \eprint{arXiv:1208.2069}.

\bibitem[{\citenamefont{Dickman et~al.}(1998)\citenamefont{Dickman, Vespignani,
  and Zapperi}}]{DickmanVespignaniZapperi:1998}
\bibinfo{author}{\bibfnamefont{R.}~\bibnamefont{Dickman}},
  \bibinfo{author}{\bibfnamefont{A.}~\bibnamefont{Vespignani}},
  \bibnamefont{and} \bibinfo{author}{\bibfnamefont{S.}~\bibnamefont{Zapperi}},
  \bibinfo{journal}{Phys. Rev. E} \textbf{\bibinfo{volume}{57}},
  \bibinfo{pages}{5095} (\bibinfo{year}{1998}).

\bibitem[{\citenamefont{Pruessner and Peters}(2006)}]{PetersPruessner:2006}
\bibinfo{author}{\bibfnamefont{G.}~\bibnamefont{Pruessner}} \bibnamefont{and}
  \bibinfo{author}{\bibfnamefont{O.}~\bibnamefont{Peters}},
  \bibinfo{journal}{Phys. Rev. E} \textbf{\bibinfo{volume}{73}},
  \bibinfo{eid}{025106(R)} (pages~\bibinfo{numpages}{4})
  (\bibinfo{year}{2006}), \eprint{arXiv:cond-mat/0411709}.

\bibitem[{\citenamefont{Fey et~al.}(2010)\citenamefont{Fey, Levine, and
  Wilson}}]{FeyLevineWilson:2010b}
\bibinfo{author}{\bibfnamefont{A.}~\bibnamefont{Fey}},
  \bibinfo{author}{\bibfnamefont{L.}~\bibnamefont{Levine}}, \bibnamefont{and}
  \bibinfo{author}{\bibfnamefont{D.~B.} \bibnamefont{Wilson}},
  \bibinfo{journal}{Phys. Rev. Lett.} \textbf{\bibinfo{volume}{104}},
  \bibinfo{eid}{145703} (pages~\bibinfo{numpages}{4}) (\bibinfo{year}{2010}),
  \eprint{arXiv:0912.3206v3}.

\bibitem[{\citenamefont{Athreya and Ney}(2012)}]{athreya2012branching}
\bibinfo{author}{\bibfnamefont{K.~B.} \bibnamefont{Athreya}} \bibnamefont{and}
  \bibinfo{author}{\bibfnamefont{P.~E.} \bibnamefont{Ney}},
  \emph{\bibinfo{title}{Branching processes}}, vol. \bibinfo{volume}{196}
  (\bibinfo{publisher}{Springer-Verlag}, \bibinfo{address}{Berlin, Germany},
  \bibinfo{year}{2012}).

\bibitem[{\citenamefont{Bollob{\'a}s and B{\'e}la}(2001)}]{bollobas2001random}
\bibinfo{author}{\bibfnamefont{B.}~\bibnamefont{Bollob{\'a}s}}
  \bibnamefont{and} \bibinfo{author}{\bibfnamefont{B.}~\bibnamefont{B{\'e}la}},
  \emph{\bibinfo{title}{Random graphs}}, \bibinfo{number}{73}
  (\bibinfo{publisher}{Cambridge university press}, \bibinfo{year}{2001}).

\bibitem[{\citenamefont{Vespignani et~al.}(2000)\citenamefont{Vespignani,
  Dickman, Mu{\~n}oz, and Zapperi}}]{VespignaniETAL:2000}
\bibinfo{author}{\bibfnamefont{A.}~\bibnamefont{Vespignani}},
  \bibinfo{author}{\bibfnamefont{R.}~\bibnamefont{Dickman}},
  \bibinfo{author}{\bibfnamefont{M.~A.} \bibnamefont{Mu{\~n}oz}},
  \bibnamefont{and} \bibinfo{author}{\bibfnamefont{S.}~\bibnamefont{Zapperi}},
  \bibinfo{journal}{Phys. Rev. E} \textbf{\bibinfo{volume}{62}},
  \bibinfo{pages}{4564} (\bibinfo{year}{2000}),
  \eprint{arXiv:cond-mat/0003285}.

\bibitem[{\citenamefont{Pietronero et~al.}(1994)\citenamefont{Pietronero,
  Vespignani, and Zapperi}}]{PietroneroVespignaniZapperi:1994}
\bibinfo{author}{\bibfnamefont{L.}~\bibnamefont{Pietronero}},
  \bibinfo{author}{\bibfnamefont{A.}~\bibnamefont{Vespignani}},
  \bibnamefont{and} \bibinfo{author}{\bibfnamefont{S.}~\bibnamefont{Zapperi}},
  \bibinfo{journal}{Phys. Rev. Lett.} \textbf{\bibinfo{volume}{72}},
  \bibinfo{pages}{1690} (\bibinfo{year}{1994}).

\bibitem[{\citenamefont{Lin}(2010)}]{lin2010renormalization}
\bibinfo{author}{\bibfnamefont{C.-Y.} \bibnamefont{Lin}},
  \bibinfo{journal}{Phys. Rev. E} \textbf{\bibinfo{volume}{81}},
  \bibinfo{pages}{021112} (\bibinfo{year}{2010}).

\bibitem[{\citenamefont{Dickman et~al.}(2002)\citenamefont{Dickman, Tom{\'e},
  and de~Oliveira}}]{dickman2002sandpiles}
\bibinfo{author}{\bibfnamefont{R.}~\bibnamefont{Dickman}},
  \bibinfo{author}{\bibfnamefont{T.}~\bibnamefont{Tom{\'e}}}, \bibnamefont{and}
  \bibinfo{author}{\bibfnamefont{M.~J.} \bibnamefont{de~Oliveira}},
  \bibinfo{journal}{Phys. Rev. E} \textbf{\bibinfo{volume}{66}},
  \bibinfo{eid}{016111} (pages~\bibinfo{numpages}{8}) (\bibinfo{year}{2002}).

\bibitem[{\citenamefont{Gleeson}(2011)}]{gleeson2011high}
\bibinfo{author}{\bibfnamefont{J.~P.} \bibnamefont{Gleeson}},
  \bibinfo{journal}{Physical Review Letters} \textbf{\bibinfo{volume}{107}},
  \bibinfo{pages}{068701} (\bibinfo{year}{2011}).

\bibitem[{\citenamefont{Gleeson}(2013)}]{gleeson2013binary}
\bibinfo{author}{\bibfnamefont{J.~P.} \bibnamefont{Gleeson}},
  \bibinfo{journal}{Physical Review X} \textbf{\bibinfo{volume}{3}},
  \bibinfo{pages}{021004} (\bibinfo{year}{2013}).

\bibitem[{\citenamefont{Rossi et~al.}(2000{\natexlab{b}})\citenamefont{Rossi,
  Pastor-Satorras, and Vespignani}}]{Rossi:2000}
\bibinfo{author}{\bibfnamefont{M.}~\bibnamefont{Rossi}},
  \bibinfo{author}{\bibfnamefont{R.}~\bibnamefont{Pastor-Satorras}},
  \bibnamefont{and}
  \bibinfo{author}{\bibfnamefont{A.}~\bibnamefont{Vespignani}},
  \bibinfo{journal}{Physical review letters} \textbf{\bibinfo{volume}{85}},
  \bibinfo{pages}{1803} (\bibinfo{year}{2000}{\natexlab{b}}).

\bibitem[{\citenamefont{Lindvall}(2002)}]{lindvall2002lectures}
\bibinfo{author}{\bibfnamefont{T.}~\bibnamefont{Lindvall}},
  \emph{\bibinfo{title}{Lectures on the coupling method}}
  (\bibinfo{publisher}{Courier Corporation}, \bibinfo{year}{2002}).

\bibitem[{\citenamefont{Levin and Peres}(2017)}]{levin2017markov}
\bibinfo{author}{\bibfnamefont{D.~A.} \bibnamefont{Levin}} \bibnamefont{and}
  \bibinfo{author}{\bibfnamefont{Y.}~\bibnamefont{Peres}},
  \emph{\bibinfo{title}{Markov chains and mixing times}}, vol.
  \bibinfo{volume}{107} (\bibinfo{publisher}{American Mathematical Soc.},
  \bibinfo{year}{2017}).

\end{thebibliography}

\end{document}